\documentclass[12pt]{iopart}
\pdfoutput=1

\usepackage{iopams}
\usepackage{amsthm}
\usepackage{dsfont}
\usepackage{graphicx}

\newcommand{\ve}[1]{\mathbf{#1}}
\newcommand{\bra}[1]{\left\langle #1 \right|}
\newcommand{\ket}[1]{\left| #1 \right\rangle}

\def\({\left(}
\def\){\right)}

\newtheorem{Prop}{Proposition}

\begin{document}
\title[RRDPS based on signal disturbance]{A security proof of the round-robin differential phase shift quantum key distribution protocol based on the signal disturbance}
\author{
 Toshihiko Sasaki$^{1}$ and Masato Koashi$^1$
}
\address{
$^1$Photon Science Center, School of Engineering, University of Tokyo, 7-3-1 Hongo, Bunkyo-ku, Tokyo 113-8656, Japan\\
}
\ead{sasaki@qi.t.u-tokyo.ac.jp}
\begin{abstract}
 The round-robin differential phase shift (RRDPS) quantum key distribution (QKD) protocol is a unique quantum key distribution protocol
 whose security has not been understood through an information-disturbance trade-off relation,
 and a sufficient amount of privacy amplification was given independently of signal disturbance.
 Here, we discuss the security  of the RRDPS protocol in the asymptotic regime
 when a good estimate of the bit error rate is available as a measure of signal disturbance.
 The uniqueness of the RRDPS protocol shows up as a peculiar form of information-disturbance trade-off curve.
 When the length of a block of pulses used for encoding and the signal disturbance are both small, it provides a significantly better key rate than that from the original security proof.
 On the other hand, when the block length is large, the use of the signal disturbance makes little improvement in the key rate.
 Our analysis will bridge a gap between the RRDPS protocol and the conventional QKD protocols.

\end{abstract}

%\pacs{03.65.Ud, 03.67.-a, 03.67.Lx.}

\maketitle
\section{Introduction}
Quantum key distribution (QKD) \cite{Bennett1984,Ekert1991,Bennett1992b,Brus1998,Scarani2004,Stucki2005,Inoue2003,Grosshans2002}  is an intensively studied field in quantum information theory 
because of its practical utility and feasibility.
It can provide an information-theoretic security in which it is assumed that an eavesdropper (Eve) is only limited by the law of nature, or quantum mechanics.
The widely understood idea  about why QKD can guarantee the security is that any attempt to read information encoded on a quantum state
causes a disturbance \cite{Heisenberg1927,Ozawa2004}.
If we base the security on this idea, we have to monitor signal disturbances to detect Eve's intervention.
Recently a new path to guarantee an information-theoretic security was proposed \cite{Sasaki2014}.
This implementation is called round-robin differential phase shift QKD, or RRDPS-QKD.
The main difference is that it does not need any monitoring of signal disturbance to ensure the security.
Although this approach simplifies the security proof and reduces overheads caused by the finite-key effect,
it is an interesting question whether we can also understand the RRDPS protocol in the standard way,
namely, with the help of monitoring the disturbance in the transmitted signals.
In this paper, we prove the security of the RRDPS protocol against general attacks with asymptotic key lengths
when it is augmented by a procedure of monitoring the bit error rates.
This analysis will gives us a deeper understanding on how the RRDPS protocol is different from or similar to the conventional QKD protocols.
It also provides a better key rate for small block sizes
and implies the optimality of the original security proof for large block sizes.

\section{Protocol}

\begin{figure}
 \begin{center}
	\includegraphics[width=12cm]{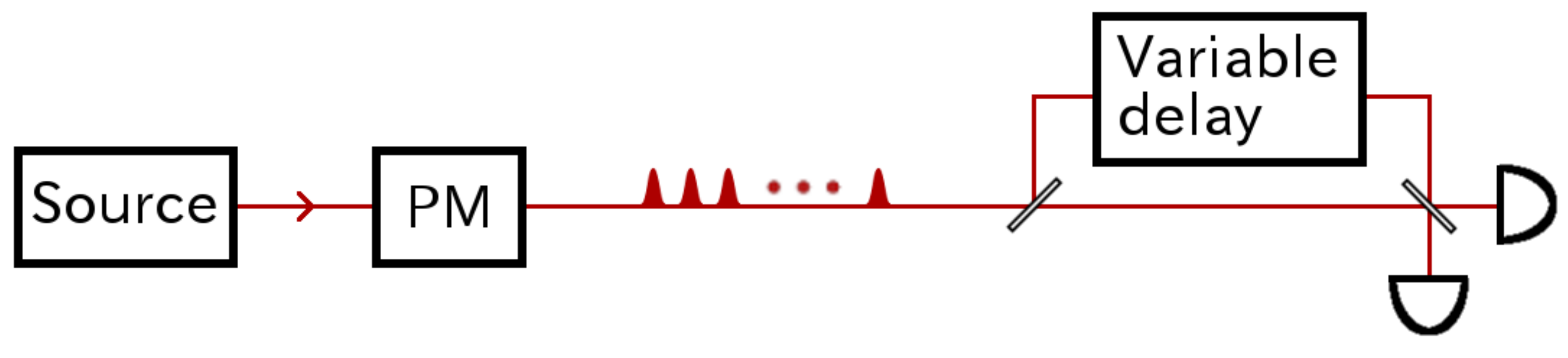}
	\caption{A schematic of a setup for the RRDPS protocol.
	Alice sends out a train of $L$ pulses after randomly applying a phase shift, either 0 or $\pi$ to each pulse with a common random phase offset $\delta$
	by use of the phase modulator (PM).
	Bob uses the variable-delay interferometer to measure a relative phase between a randomly chosen pair.
	The amount of the delay is chosen randomly  from $1$ to $L-1$ times the interval of the adjacent pulses.
	}
\label{protocol-setup}
 \end{center}
\end{figure}

We describe a setup of the RRDPS protocol.
The sender Alice prepares an optical pulse train consisting of $L$ pulses.
The optical phase of each pulse is randomly modulated by $0$ or $\pi$.
We call these $L$ pulses a block and $L$  the block size.
Alice applies a common random phase shift $\delta$ to all pulses in the block.
The receiver Bob uses the variable-delay interferometer to measure a relative phase between a randomly chosen pair \cite{Honjo2006}.
A schematic of this interferometer is illustrated in \fref{protocol-setup}.
The delay of this interferometer is randomly chosen from $1$ to $L-1$ times the interval of the adjacent pulses.
Bob uses photon-number-resolving photon detectors to measure the relative phase of two pulses in the block.
At the same time, Bob makes sure the total number of photons in the received block is unity, and otherwise declares a failure of detection.
The details of this check process and its variants are given in \ref{appendix-one-photon}.
If the detection has succeeded, Bob sets his pre-sifted key bit as 0 (1) if the relative phase of the detected pair is 0 ($\pi$).
From the timing of the detection and the amount of the chosen delay, Bob can determine the indices of the detected pair.
He discloses the pair of indices over by an authenticated public channel.
Alice sets her pre-sifted key bit by calculating the relative phase of the announced pair according to her record of the modulations.
We refer to the above procedure involving transmission of one block of pulses as a round.
After repeating a predetermined number ($N$) of rounds, Alice and Bob announce their pre-sifted key bits of randomly sampled rounds to determine the bit error rate and remove these key bits from each of the pre-sifted keys to define their sifted keys.
We call an un-sampled round, in which a sifted key bit is produced, as a sifted round.
Alice sends an encrypted syndrome of her sifted key to Bob.
Bob reconciles his key to Alice's key by use of this syndrome.
Then, they perform privacy amplification to obtain a correct and secure key with small imperfection.

As for the imperfections of the detectors, we assume that a non-unit quantum efficiency of a detector is modeled by a linear absorber
followed by a perfect detector with unit quantum efficiency. 
We further assume that the all detectors have the same quantum efficiency.
Then the inefficiency can be modeled by a common linear absorber placed in front
of the interferometer, which can be included in the quantum channel without compromising the security.

As for Alice's light source, we assume that the probability that it emits an $L$-pulse block containing more photons than a fixed threshold photon number $\nu_{\mathrm{th}}$
is at most $e_{\mathrm{src}}$.
This condition can be certified in an offline test.
In the security proof, we do not assume that the photon number distribution of the light source is a Poisson distribution.

\section{Security}
We will prove the security of the RRDPS protocol by use of the Shor-Preskil argument extended in \cite{Koashi2009} and the technique in \cite{Tamaki2012}.
In this argument, a sufficient amount of the privacy amplification to guarantee the security of Alice's key is determined by estimating how well the results of a complementary measurement can be predicted.

In order to define a complementary measurement for this protocol, we construct an alternative procedure which is equivalent to the actual setup as follows.
In the actual setup, Alice generates a random bit sequence $s_1,s_2,\cdots,s_L$ and emits the corresponding optical state
$
 e^{i\delta\sum_{k}\hat{n}_k} \(\bigotimes_{k=1}^L(-1)^{s_k \hat{n}_k}\)\ket{\Psi},
$
where $\hat{n}_k$ is the number operator of the $k$-th pulse and $\ket{\Psi}$ is the state of the $L$-pulse train.
Since the common phase shift $\delta$ is chosen randomly and is never referred to,
we can identify the emitted state as the classical mixture of the states projected to the subspaces with fixed total photon number.
The same distribution of the sequence $s_1,s_2,\cdots,s_L$ and the corresponding emitted state is obtained if Alice uses $L$ auxiliary qubits (denoted by $(A,1),\cdots, (A,L)$)
and a classical register (denoted by $c$) recording the total photon number to prepare the state
\begin{eqnarray}
\label{state-prepare}
2^{-L/2}\sum_{\nu=0}^{\infty} \ket{\nu}_{c} \hat{\pi}_{\nu}\bigotimes_{k=1}^L \(\sum_{s_k\in\{0,1\}}\ket{s_k}_{A,k}(-1)^{s_k\hat{n}_k}\)\ket{\Psi},
\end{eqnarray}
and then measures the auxiliary qubits on the $\{\ket{0}_{A,k},\ket{1}_{A,k}\}$ bases to determine $s_k$.
Here, $\hat{\pi}_\nu$ is the projector onto the state whose number of photons in the block is $\nu$.
We call the $\{\ket{0}_{A,k},\ket{1}_{A,k}\}$ basis as $Z$ basis of the $k$-th qubit.
We also replace Alice's procedure of determining her pre-sifted key bit.
In the actual protocol, after hearing Bob's announcement of a pair $\{k,l\}$,
Alice determines her pre-sifted key bit by computing classical XOR $s_k\oplus s_l$.
An equivalent procedure can be constructed as follows.
Instead of measuring each of the auxiliary qubits on $Z$ basis, we apply a CNOT gate to the $k$-th and $l$-th qubits
and then measure the target qubit on the $Z$ basis.
%We call this target qubit after the CNOT gate operation as Alice's final qubit.
Although the above description of the equivalent procedure is ambiguous about the assignment of control and target qubits,
it will be clarified later.

We define the complementary measurement for this protocol as the $X$-basis measurement on the Alice's target qubit, where
$X$-basis measurement is defined as $\{\ket{+},\ket{-}\}$ basis measurement, and
$\ket{\pm}$ is defined as $\(\ket{0}\pm\ket{1}\)/\sqrt{2}$.
We also define the $\{\ket{+}_{A,k},\ket{-}_{A,k}\}$ basis as $X$ basis of the $k$-th qubit.
Note that the state $\ket{+}_{A,k}$ ($\ket{-}_{A,k}$) corresponds to the case where the $k$-th pulse contains an even (odd) number of photons
(see \eref{security-phi-pn}).
If no photons are emitted in a block, each auxiliary qubit is left in state $\ket{+}_{A,k}$,
and hence the state of the target qubit is always $\ket{+}$.
This means that the result of the complementary measurement is predicted with certainty,
and it is compatible with the fact that no information should have leaked on Alice's sifted key when no photons were emitted.
Based on this observation, we define the occurrence of  a ``phase error'' to be the case where the complementary measurement results in state $\ket{-}$.

The use of a laser as a light source implies that there is no deterministic bound on the total photon number $\nu$.
To handle this, we use the tagging idea \cite{Gottesman2004}.
In the equivalent procedure defined above, \eref{state-prepare} implies that Alice can measure the register $c$ to learn the emitted total photon number $\nu$ in each round.
We define a round to be tagged if $\nu>\nu_{\mathrm{th}}$.
We pessimistically assume that the full information is leaked for the tagged rounds.
Let $\delta_{\mathrm{tag}}$ be an upper bound on the ratio of the number of tagged and detected rounds against the total number of detected rounds.
Suppose that there is an upper bound $\bar{e}^{(\mathrm{ph})}_{\mathrm{unt}}(\leq 1/2)$ on the phase error rate in the untagged and detected rounds,
namely, the number of phase error in the untagged and detected rounds against the total number of the untagged and detected rounds.
According to \cite{Koashi2009,Gottesman2004}, the asymptotic key rate per block is then given by
\begin{equation}
 \label{keyrate-tmp-formula-01}
 G/N = Q\( 1 - \mathrm{EC} - \mathrm{PA}\),
\end{equation}
\begin{equation}
  \label{keyrate-tmp-formula-02}
 \mathrm{EC} = f_{\mathrm{EC}}h(e),
\end{equation}
\begin{equation}
 \label{keyrate-tmp-formula-03}
 \mathrm{PA} = \delta_{\mathrm{tag}}+ \(1-\delta_{\mathrm{tag}}\)h(\bar{e}^{(\mathrm{ph})}_{\mathrm{unt}}),
\end{equation}
where $N$ is the total number of the emitted blocks, $Q$ is the detection rate per block, $\mathrm{EC}$ is the cost for the error correction,
$\mathrm{PA}$ is the cost for the privacy amplification, $h(x)$ is the binary entropy function defined as $-x\log_2 x -(1-x)\log_2 (1-x)$,
$e$ is the bit error rate, and $f_{\mathrm{EC}}$ is determined from the efficiency of the bit error correction.
Our main task now is to determine the bound $\bar{e}^{(\mathrm{ph})}_{\mathrm{unt}}$ under the condition that the observed bit error rate is $e$ and that the emitted photon number is at most $\nu_{\mathrm{th}}$.

First, we focus on how the pair $\{k,l\}$ is determined.
Since Bob announces the pair only when he has received only one photon in a block,
the positive-operator valued measure (POVM) element $\hat{P}'_{\{k,l\},s_B}$of the actual measurement that determines the unordered pair $\{k,l\}$  and the bit $s_B$ corresponding to the phase difference
is written as
\begin{equation}
 \hat{P}'_{\{k,l\},s_B} = \frac{1}{2(L-1)}\hat{P}\(\frac{1}{\sqrt{2}}\(\ket{k}_B+ (-1)^{s_B}\ket{l}_B\)\),
\end{equation}
where $\hat{P}(\ket{\psi})$ is defined as $\ket{\psi}\bra{\psi}$, and $\ket{k}$ refers to the state of the $L$ pulses containing one photon in the $k$-th pulse  and no other photons.
Note that the success of detection corresponds to the operator
\begin{equation}
 \sum_{\{k,l\}\in \mathfrak{P},s_B\in\{0,1\}} \hat{P}'_{\{k,l\},s_B} = \frac{1}{2}\sum_{k\in\{1,\cdots,L\}}\hat{P}\({\ket{k}_B}\),
\end{equation}
where we define $\mathfrak{P}$ as $\{\{k,l\}\mid k<l, k,l,\in \{1,\cdots,L\}\}$.
This equation means that as long as the received block contains one photon in total, Bob's detection succeeds with probability 1/2 regardless of its state.
We can thus rewrite the actual measurement equivalently to the three-step measurement procedure defined as
\begin{enumerate}
 \item \label{step1}Bob confirms that there is one photon in the received block via an ideal projection (quantum non-demolition) measurement.
			 If not, the measurement procedure ends with a failure of detection.
 \item \label{step2}With a probability of 1/2, Bob ends the measurement procedure with a failure of detection.
 \item \label{step3}Reaching this step implies that the success of detection has been assured.
			 Bob performs the measurement defined by $\hat{P}_{\{k,l\},s_B}(:=2\hat{P}'_{\{k,l\},s_B})$
			 on the $L$ pulses to determine an outcome ($\{k,l\},s_B$).
			 He announces the unordered pair $\{k,l\}$.
\end{enumerate}

In order to assess the security of Alice's final key, we introduce a virtual protocol
in which Bob decides, prior to the step (\ref{step3}), whether the round is to be sampled or not.
Further, when it is not to be sampled, he forgoes the step (\ref{step3}) of the three-step measurement procedure and instead announces an ordered pair $(k,l) (k\neq l)$ via a measurement described by the POVM elements
\begin{equation}
 \label{def-ordered-P}
 \hat{P}_{(k,l)} = \frac{1}{(L-1)}\hat{P}\(\ket{k}_B\).
\end{equation}
This measurement is understood as it determines the index $k$ of the pulse containing the single photon
and randomly chooses the other index $l$.
From the viewpoint of determining the unordered pair $\{k,l\}$, this virtual measurement is equivalent to the actual measurement because
\begin{equation}
 \hat{P}_{(k,l)} + \hat{P}_{(l,k)} = \hat{P}_{\{k,l\},0}+\hat{P}_{\{k,l\},1}.
\end{equation}
This means that any attack strategy by Eve can be applied to the virtual protocol
and leads to the same final state over Alice's final key and Eve's quantum system as in the actual protocol.
Therefore, the security of the virtual protocol implies that of the actual one.
Now we remove the ambiguity about the assignment of control and target qubits for the CNOT gate.
We choose the index of the control qubit to be $k$ when the ordered pair $(k,l)$ was announced.
We can then write the POVM element $\hat{e}^{(\mathrm{ph})}_{(k,l)}$
corresponding to the announcement of $(k,l)$ and occurrence of a phase error (finding the target qubit in state $\ket{-}$) as
\begin{equation}
\label{security-odd-even}
 \hat{e}^{(\mathrm{ph})}_{(k,l)} 
= \hat{P}(\ket{-}_{A,l}) \hat{P}_{(k,l)},
\end{equation}
where we have used invariance of $P(\ket{-}_{A,l})$ under the CNOT gate.
We also define the POVM element $\hat{e}^{(\mathrm{ph})}$ corresponding to the occurrence of a phase error as
\begin{equation}
 \hat{e}^{(\mathrm{ph})} = \sum_{\{k,l\}\in \mathfrak{P} } \hat{e}^{(\mathrm{ph})}_{(k,l)}+\hat{e}^{(\mathrm{ph})}_{(l,k)}.
\end{equation}

In order to bound
the phase error rate $e^{(\mathrm{ph})}_{\mathrm{unt}}$ in the untagged sifted rounds in the virtual protocol,
we utilize the bit error rate in the sample, which is determined commonly in the actual and the virtual protocol.
We also use the condition on the total photon number in an untagged round.
In accordance with the definition of $\hat{e}^{\mathrm{ph}}$, we regard the occurrence of a bit error as an outcome of a joint measurement on Alice's $L$ qubits and Bob's $L$ pulses prior to the step (\ref{step3}).
The corresponding POVM element $\hat{e}$ is given by
\begin{equation}
 \hat{e} =  \sum_{\{k,l\}\in \mathfrak{P},s,s_B\in\{0,1\}} \hat{P}\(\ket{s}_{A,k}\ket{s\oplus s_B\oplus 1}_{A,l}\) \hat{P}_{\{k,l\},s_B}
\end{equation}
As for the total photon number, \eref{state-prepare} dictates that the state of Alice's $L$ qubits is confined in a subspace $\mathcal{H}_{A,\nu}$
when the emitted total photon number is $\nu$.
Let $\hat{P}^{(\nu)}$ be the projector to $\mathcal{H}_{A,\nu}$.
An asymptotic relation between the phase error rate $e^{(\mathrm{ph})}_{\mathrm{unt}}$ in the untagged sifted rounds and the bit error rate $e_{\mathrm{unt}}$ in the untagged sampled rounds is then derived
by considering the maximum eigenvalue $\Omega(\nu,\lambda)$ of the operator $\hat{P}^{(\nu)}(\hat{e}^{(\mathrm{ph})} - \lambda \hat{e})\hat{P}^{(\nu)}$
for $\nu\leq \nu_{\mathrm{th}}$ and a positive number $\lambda$. 
The detailed calculation given in \ref{appendix-main-theorem} leads to the following.
In the asymptotic limit, we have
 \begin{equation}
	\label{prop-main-first-ineq}
	  e^{(\mathrm{ph})}_{\mathrm{unt}} \leq F\(\nu_{\mathrm{th}},e_{\mathrm{unt}}\),
 \end{equation}
 where $F(\nu,e)$ is defined as
 \begin{equation}
	\label{prop-main-second-ineq}
	F(\nu,e) := \inf_{\lambda\geq 0} \(\lambda e + \Omega(\nu,\lambda)\),
 \end{equation}
 \begin{equation}
		\label{prop-main-third-ineq}
	\Omega(\nu,\lambda) := \max\(\Omega_-(\nu,\lambda),\Omega_+(\nu,\lambda)\),
 \end{equation}
 \begin{equation}
		\label{prop-main-fourth-ineq}
 \Omega_-(\nu,\lambda) := \frac{\nu-1}{L-1} - \frac{L\lambda+2}{4(L-1)}\( 1- \sqrt{ 1- \frac{8(\nu-1)\lambda}{(L\lambda+2)^2}}\),
 \end{equation}
 and
 \begin{equation}
		\label{prop-main-fifth-ineq}
 \Omega_+(\nu, \lambda) := \frac{\nu}{L-1} - \lambda\frac{L-1-\nu}{2(L-1)}.
 \end{equation}

 \begin{figure}[t]
	 \begin{center}
		\includegraphics[width=7cm]{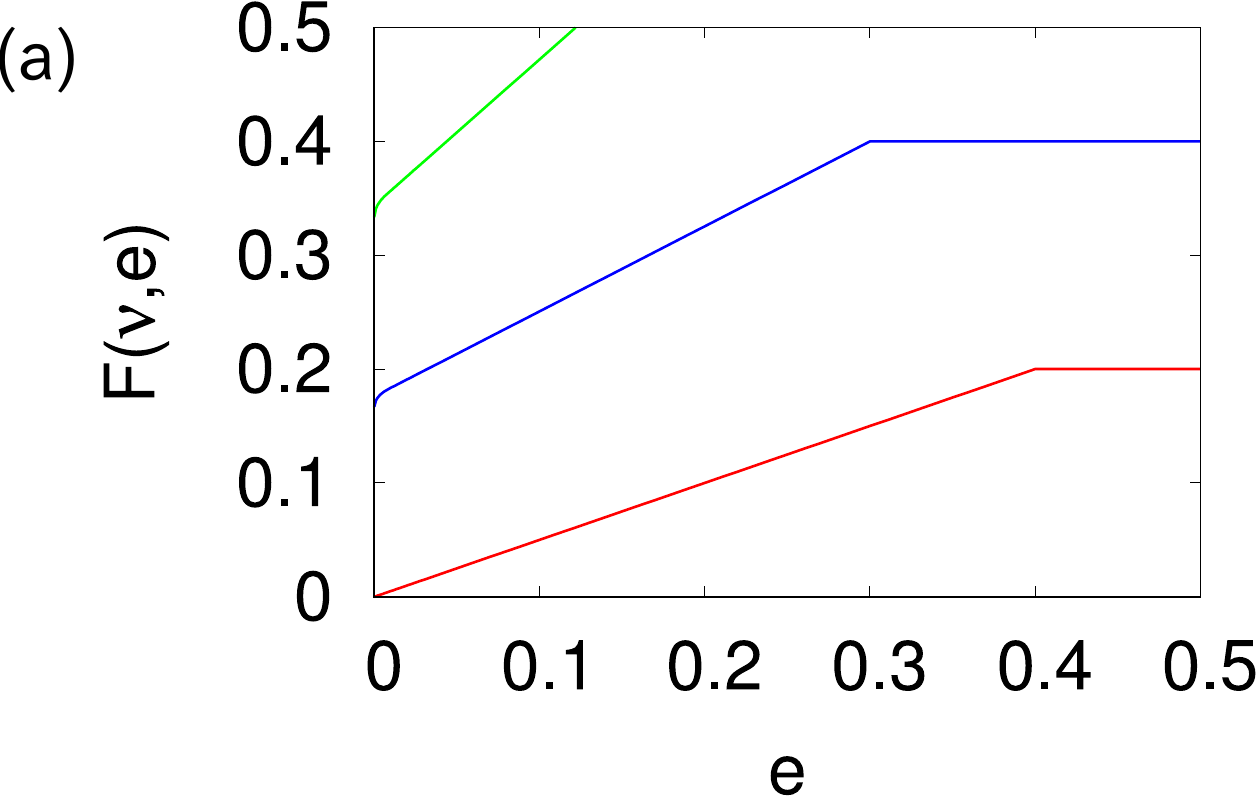}
		\includegraphics[width=7cm]{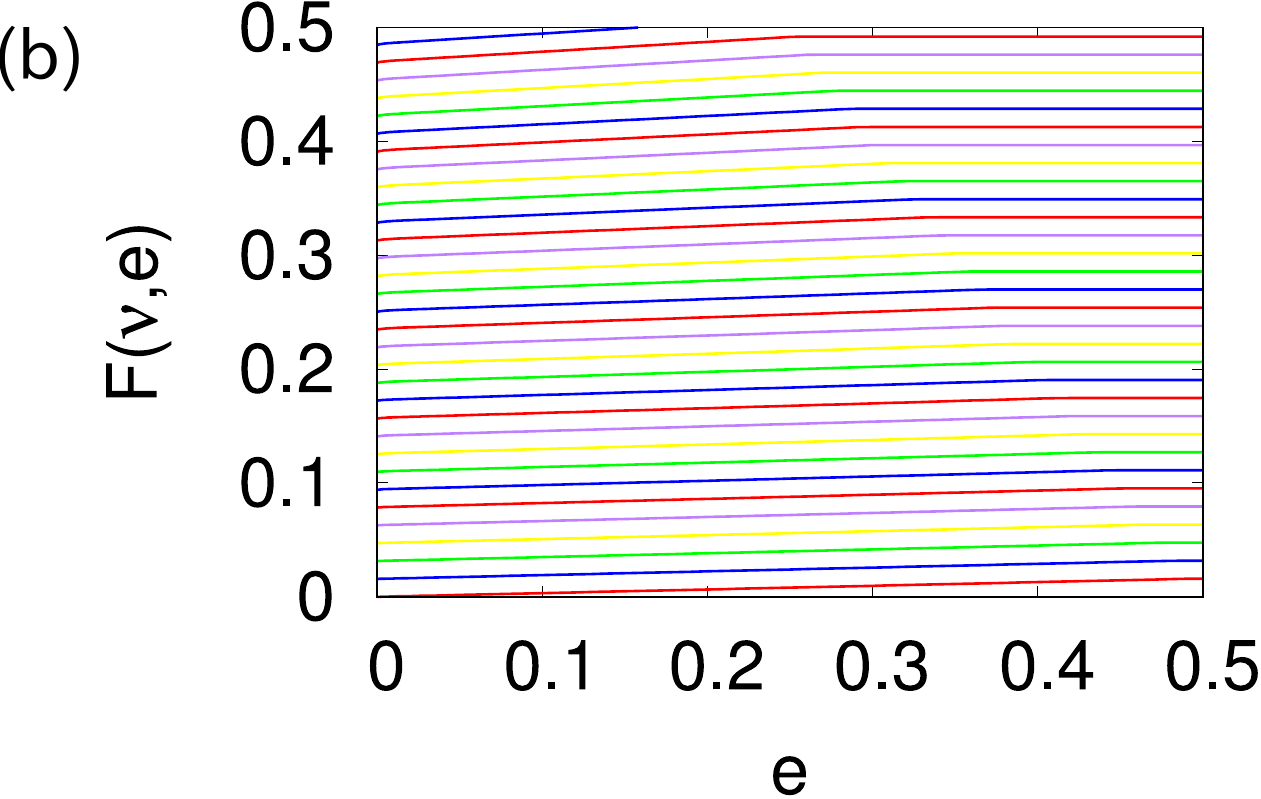}
	 \caption{The graph of the function $F(\nu,e)$,
	 which also implicitly depends on the block size $L$.
	 (a) For $L=6$, we vary $\nu$ from 1 to 3.
	 (b) For $L=64$, we vary $\nu$ from 1 to 32.
	 The value of $F(\nu,e)$ increases as $\nu$ becomes larger.
	 }
\label{fig-e-eph}
	 \end{center}
\end{figure}

We illustrate the function $F(\nu,e)$ for the cases $L=6$ and $L=64$ in \fref{fig-e-eph}.
As is seen from the figure, each curve for $F(\nu,e)$ with fixed $\nu$ can be well approximated
by two straight segments, given by
\begin{equation}
 F(\nu,e)\sim	\cases{
	\frac{\nu-1}{L} \(1-\frac{e}{e^\ast(\nu)}\)+\frac{\nu}{L-1}\frac{e}{e^\ast(\nu)} & $0\leq e < e^\ast(\nu)$\\
 \frac{\nu}{L-1} & $e \geq e^\ast(\nu)$\\
 }
\end{equation}
where $e^\ast(\nu)$ is defined as $(L-1-\nu)/(2(L-1))$.
The function $F(\nu,e)$ takes a constant value when $e$ is large.
The value of $\nu/(L-1)$ reproduces the result of the original security proof \cite{Sasaki2014}.
There, the bound was derived by an argument which relies neither on the state of the pulses after Eve's intervention nor correlations between Alice's and Bob's systems.
Roughly speaking, Eve cannot learn most of the phases of the $L$ pulses emitted from Alice since those pulses are weak.
Moreover, as seen in \eref{def-ordered-P}, one of the announced indices is uniformly random no matter what state Eve sends to Bob. These two properties prevent Eve from predicting the phase difference in the announced pair.
When $e$ is small, the bound $F(\nu,e)$ becomes lower as $e$ becomes smaller,
which is a typical information-disturbance trade-off.
Qualitatively, the origin of the trade-off is the same as the original DPS protocol \cite{Inoue2003}.
If Eve reads the relative phase between two successive weak pulses, she disturbs the relative phases of these pulses to the other pulses,
increasing the bit error rate.
More quantitatively, however, it was also shown  \cite{Tsurumaru2007,Tamaki2012} that
Eve can considerably suppress the increase in the bit error rate by gradually modulating the amplitudes of a sequence of pulses in the DPS protocol.
Since the relative phases of two distant pulses in this strategy are highly disturbed, it causes a high bit error rate in the RRDPS protocol.

The remaining problem is to bound $\delta_{\mathrm{tag}}$ and $e_{\mathrm{unt}}$.
From the definition of $e_{\mathrm{src}}$, the expected total number of the tagged rounds is at most $N e_{\mathrm{src}}$.
If we recall Bob's three-step measurement procedure, we see that half of the rounds lead to detection failure at step (\ref{step2}).
In the asymptotic limit, it means 
 \begin{equation}
	\delta_{\mathrm{tag}} \leq \frac{e_{\mathrm{src}}}{2Q}.
 \end{equation}
 As for $e_{\mathrm{unt}}$,
 we use the fact that a fraction $(1-\delta_{\mathrm{tag}})$ of the sampled rounds should be untagged in the asymptotic limit.
 Then, we have $e_{\mathrm{unt}}(1-\delta_{\mathrm{tag}})\leq e $, namely,  
 \begin{equation}
	e_{\mathrm{unt}} \leq \frac{e}{1-\delta_{\mathrm{tag}}}.
\end{equation} 

	Combining these bounds with \eref{keyrate-tmp-formula-03}, we obtain a sufficient amount of the privacy amplification as
	\begin{equation}
	 \label{eq-PA-final}
	\mathrm{PA} = \frac{e_{\mathrm{src}}}{2Q}+\(1-\frac{e_{\mathrm{src}}}{2Q}\)h\(F\(\nu_{\mathrm{th}}, \frac{e}{1-\frac{e_{\mathrm{src}}}{2Q}}\)\).
	\end{equation}
\section{Secure key rate}

\begin{figure}[t]
 \begin{center}
\includegraphics[width=7cm]{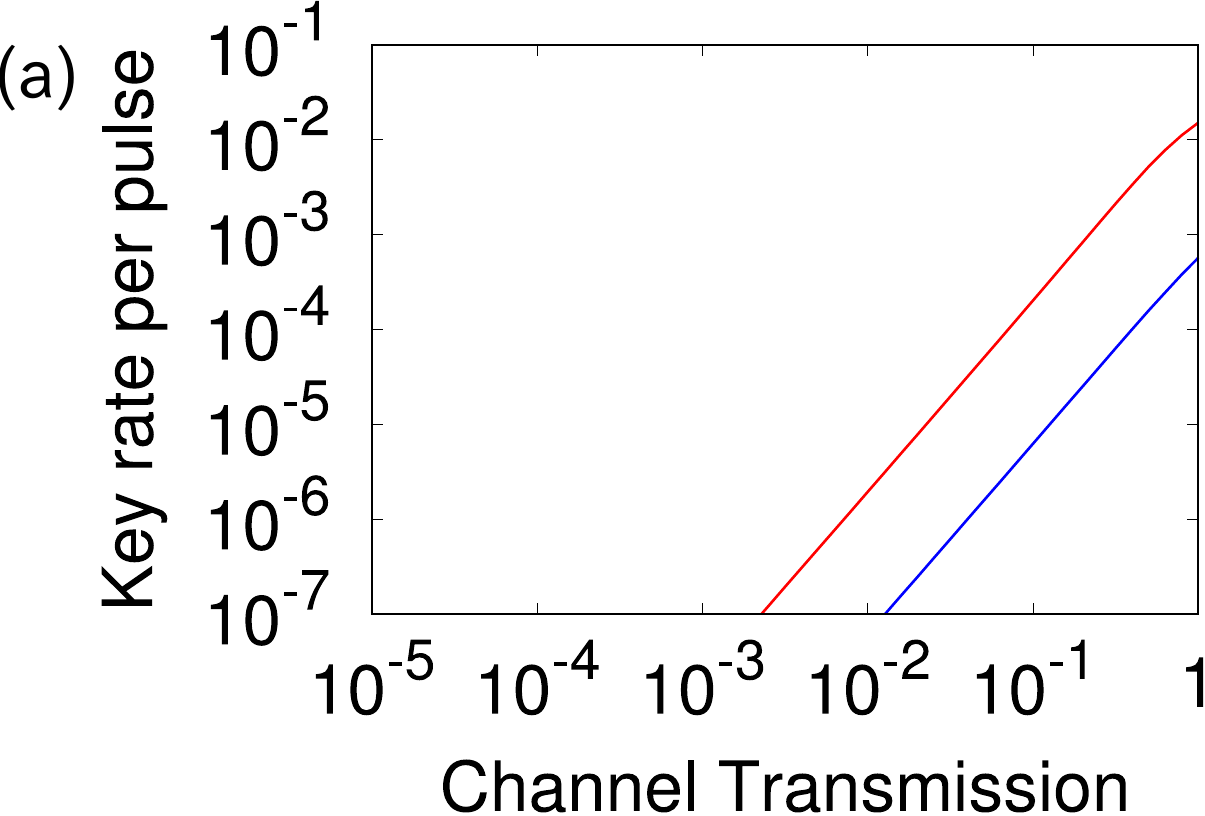}
\includegraphics[width=7cm]{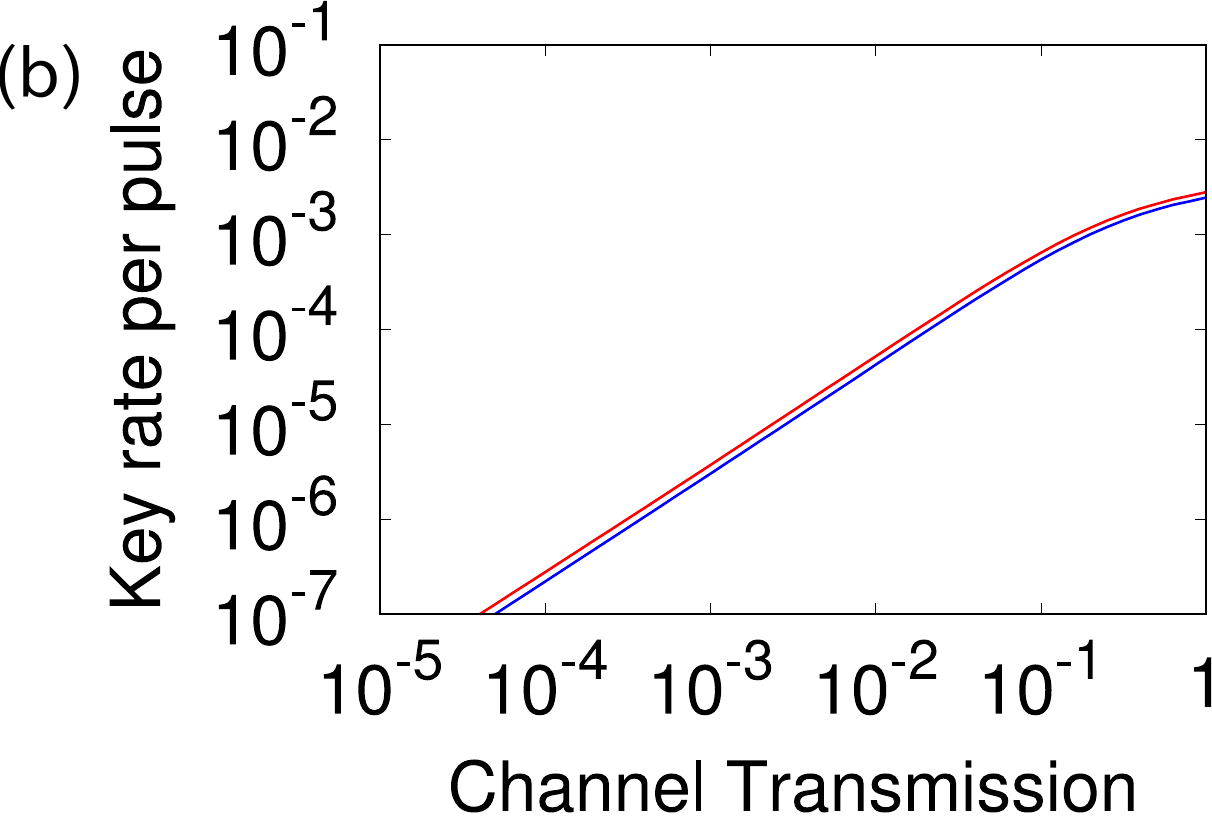}
\caption{ Asymptotic key rate per pulse $G/(LN)$ versus
overall transmission $\eta$ for (a) $L=6$ and (b) $L=64$.
We assume a constant bit error rate of $3\%$.
The upper curve: the key rate when the signal disturbance is monitored.
The lower curve: the key rate when the signal disturbance is not monitored.
}
\label{fig-key-rate-transmission}	
 \end{center}
\end{figure}

In order to examine how the performance of the RRDPS protocol is improved by use of the bit error rate, we calculate the secure key generation rate based on the following model.
We assume that the dark countings are negligible and set $e=0.03$ and $f_{\mathrm{EC}}=1.1$.
We also assume that the measurement results are the same as those calculated with the source emitting coherent states.
It means that $Q$ and $e_{\mathrm{src}}$ is written as
\begin{equation}
 Q = \frac{L\mu\eta}{2}\exp\(-\frac{L\mu\eta}{2}\)
\end{equation}
and 
\begin{equation}
 e_{\mathrm{src}} = 1-\exp(-L\mu)\sum_{\nu=0}^{\nu_{\mathrm{th}}} \frac{(L\mu)^\nu}{\nu!},
\end{equation}
where $\mu$ is the average photon number of a pulse and $\eta$ is the overall transmission rate.
In \fref{fig-key-rate-transmission}, we plot the key rate per pulse  $G/(LN)$ as a function of the overall transmission rate $\eta$ for $L=6$ and $L=64$.
We optimized the values of $\nu_{\mathrm{th}}$ and $\mu$ for each $\eta$.
We also plotted the key rate when the amount of privacy amplification is independent of the bit error rate,
which is given by \eref{eq-PA-final} with $F(\nu,e)$ replaced by $\nu_{\mathrm{th}}/(L-1)$.
These graphs confirm that we can improve the key rate by use of signal disturbance.
On the other hand, the amount of improvement decreases as the block size becomes larger.
This behavior can be ascribed to the gap between $F(\nu,0)$ and $F(\nu,0.5)$,
namely, how the ability of Eve to cause a phase error depends on whether her attack is restricted to a low bit error rate or not.
In the limit of no bit errors, what Eve can do is to leave one of the $\nu$ emitted photons untouched and pass on to Bob.
She gathers information from the remaining $\nu-1$ photons,
while she has no control over which indices will be announce by Bob
because the photon passed on to Bob is coherent over entire pulses and hence there is no correlation to the other systems.
This explains the value of $F(\nu,0)=(\nu-1)/L$.
When we remove the restriction on the bit error, Eve can use all of the $\nu$ photons to gather information.
She can also have a partial influence on the announced indices.
As seen from \eref{def-ordered-P}, she has the power of veto on one value and the range of index $l$ spans $L-1$ values.
This results in $F(\nu,0.5)=\nu/(L-1)$.
When the block size becomes large, both values behaves as $\nu/L$, leading to a small difference in Eve's ability.

\section{Summary}
In this paper, we proved the security of the RRDPS protocol against general attacks in the asymptotic regime from the standard viewpoint
that Eve's intervention is estimated from the signal disturbance.
The key ingredient of the security proof was to determine the relation between the bit error rate and the phase error rate,
where the latter represents the lack of ability to predict an observable complementary to the sifted key.
As in the other protocols such as the BB84 protocol, the phase error rate decreases as the signal disturbance approaches to zero.
On the other hand, beyond a threshold value of the bit error rate,
the phase error ceases to depend on the bit error rate,
which gives an alternative explanation for the original RRDPS protocol in terms of information-disturbance relation.
We have seen an improvement in the key rate compared to that obtained with no monitoring of the signal disturbance.
The difference is prominent for a small value of the block size $L$, whereas it decreases with $L$.
The reason for this behavior is that the leading $\Or(L^{-1})$ term of the phase error rate is the same regardless of whether the bit error is restricted or not.
This indicates that the original RRDPS protocol for a large block size is almost optimal even though it does not use the information available from monitoring disturbance.

\ack

We thank K. Tamaki for helpful discussions.
This work was funded in part by ImPACT Program of Council for Science, Technology and Innovation (Cabinet Office, Government of Japan),  
Photon Frontier Network Program (Ministry of Education, Culture, Sports, Science and Technology).

\appendix

\section{}
\label{appendix-one-photon}

	 \begin{figure}
		\begin{center}
		 \includegraphics[width=6cm]{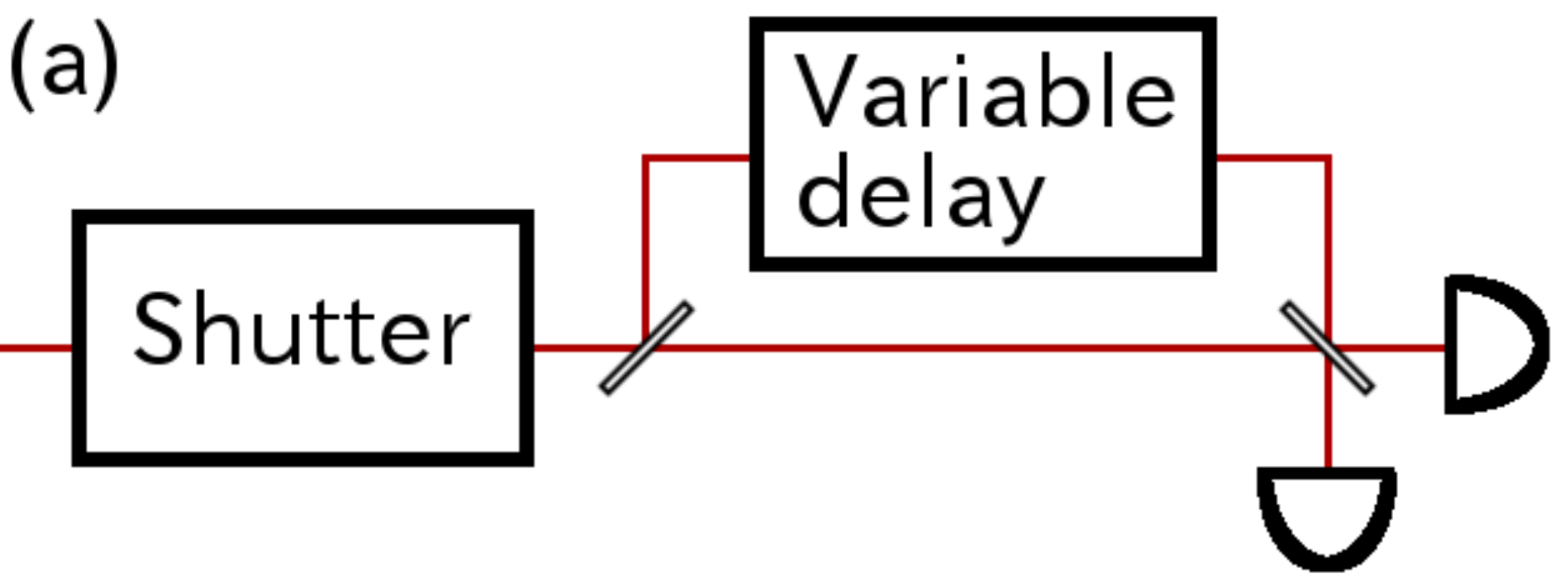}
		 \qquad 
		 \includegraphics[width=6cm]{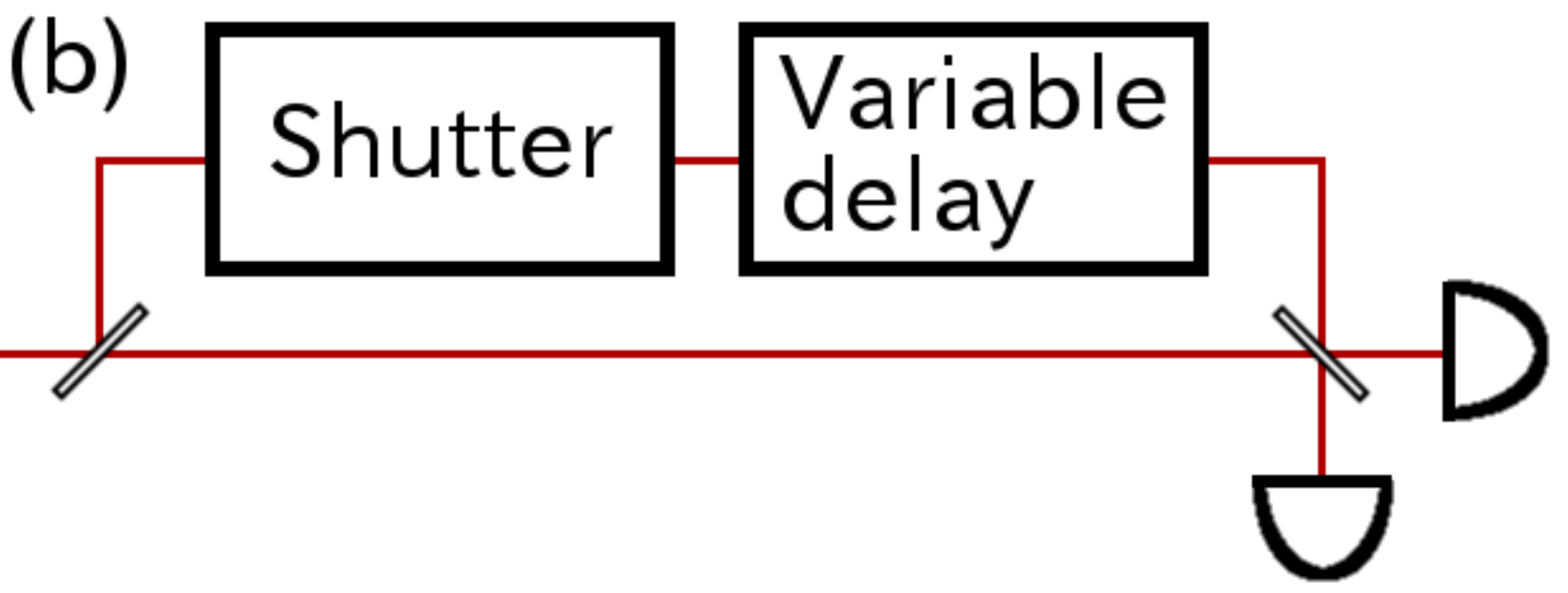}
		 \caption{(a): A setup for counting the total photon number deterministically.
		 An optical shutter is placed just before the variable-delay interferometer.
		 It is closed during the arrival timings of the idle pulses.
		 (b): A measurement setup using threshold detectors.
		 An optical shutter is placed just before the variable delay.
	When the shutter is closed for the whole block, the double count rate of this setup can bound the rate of rounds where there are two or more photons in the block received  by Bob.
	The optical shutter is also used for avoiding the mixing with the next block.
	 }
 \label{img_appB}
		\end{center}
	 \end{figure}

In order to count the total photon number by using the photon number resolving detectors after the delayed interferometer,
we need a careful implementation to avoid crosstalk between adjacent blocks.
One simple way is to define blocks such that each block is followed by $L-1$ idle pulses before the next block starts.
As in \fref{img_appB}(a), Bob places an optical shutter in front of his apparatus and close it during the arrival timings of the idle pulses.
All photons in a block are then reach a detector at one of $2L-1$ timings dedicated solely for the block.
In this scheme, the insertion of idle pulses almost halves the efficiency of the protocol.
In order to avoid the separation of blocks, Bob can prepare two copies of variable interferometers
and use a switch to redirect every other incoming block to each interferometer.

There is another way to handle this problem, which allows the use of threshold detectors (distinguishing only the vacuum from one or more photons) instead of the photon number resolving detectors, which was adopted in the demonstration \cite{Takesue2015a} of the RRDPS protocol with a passive configuration.
Instead of actually learning the photon number in each block, we use the idea of tagging, namely, the property that each multi-photon block can be tagged in principle without affecting the measurements in the protocol.
To obtain an upper bound on the number of tagged rounds, Bob places an optical shutter just before the variable delay as in \fref{img_appB}(b).
For a randomly sampled block, he throws away all the pulses in the delayed arm, and records the occurrence of coincidence detections in the entire block.
This is equivalent to a conventional setup \cite{HanburyBrown1956,Glauber1963}, which can be used for bounding the probability of multiple photons in the signal.
In this scheme, it is not necessary to separate the blocks apart
because the optical shutter can be used for avoiding the mixing with the next block
by throwing away the pulses that are to be delayed beyond the block duration.

\section{}
\label{appendix-main-theorem}
We first identify the subspace $\mathcal{H}_{A,\nu}$ of Alice's $L$ qubits and the corresponding projection $\hat{P}^{(\nu)}$.
The initial state in \eref{state-prepare} can be rewritten in terms of the $X$ basis of the qubits as
\begin{equation}
 \label{security-phi-pn}
	 \sum_{\nu=0}^{\infty} \ket{\nu}_{c} \hat{\pi}_{\nu}
	 \bigotimes_{k=1}^L \(\sum_{\pm_s\in\{+,-\}}\ket{\pm_s}_{A,k}\frac{1}{2}\(1\pm_s(-1)^{\hat{n}_k}\)\)\ket{\Psi}.
\end{equation}
The operator $\frac{1}{2}\(1\pm_s(-1)^{\hat{n}_k}\)$ is a projection onto
the subspace in which the $k$-th pulse includes even (or odd) number of photons.
Let us write the $X$-basis states of Alice's $L$ qubits as $\hat{H}^{\otimes L}\ket{\ve{a}}_A$,
where $\ve{a}$ is the array of bit values $a_{k}$, $\ket{\ve{a}}_{A}$ represents $\bigotimes_{k=1}^L\ket{a_k}_{A,k}$,
and $\hat{H}$ is the Hadamard operator.
Let  $|\ve{a}|$ be $\sum_{k=1}^L \delta_{a_k,1}$.
The correlation in \eref{security-phi-pn} implies that when the qubits are found in state  $\hat{H}^{\otimes L}\ket{\ve{a}}_A$, the total emitted photon number $\nu$ satisfies $\nu \geq |\ve{a}|$ and has the same parity as that of $|\ve{a}|$.
Hence, given $\nu$, the state of the qubits should be contained in subspace $\mathcal{H}_{A,\nu}$ defined by the projection
\begin{equation}
 \hat{P}^{(\nu)} := \sum_{|\ve{a}|\in \mathfrak{N}_{\nu}} \hat{P}\(\hat{H}^{\otimes L}\ket{\ve{a}}_A\),
\end{equation}
with $\mathfrak{N}_{\nu}:=\{\nu-2k \mid k\in \mathbb{Z}, k\geq 0, \nu-2k\geq 0 \}$.

For a non-negative number $\lambda$,
let us define $\Omega(\nu,\lambda)$ as the largest eigenvalue of the operator $\hat{P}^{(\nu)}\(\hat{e}^{(\mathrm{ph})}-\lambda\hat{e}\)\hat{P}^{(\nu)}$,
which satisfies
\begin{equation}
 \label{appendix-eph-lambda-e-ineq}
 \hat{P}^{(\nu)}\(\hat{e}^{(\mathrm{ph})}-\lambda\hat{e}\)\hat{P}^{(\nu)} \leq \Omega(\nu,\lambda)\hat{P}^{(\nu)}.
\end{equation}
Let us consider one of the untagged rounds, which means $\nu \leq \nu_{\mathrm{th}}$,
and let $\hat{\rho}_{\mathrm{unt}}$ be the state of Alice's $L$ qubits and Bob's $L$ pulses prior to the step (\ref{step3}).
From its construction, $\hat{\rho}_{\mathrm{unt}}$ satisfies $\hat{P}^{(\nu)}\hat{\rho}_{\mathrm{unt}}\hat{P}^{(\nu)}=\hat{\rho}_{\mathrm{unt}}$.
By multiplying \eref{appendix-eph-lambda-e-ineq} by $\hat{\rho}_{\nu}$ and taking the trace, we find
\begin{equation}
\Tr\(\hat{\rho}_{\mathrm{unt}} \hat{e}^{(\mathrm{ph})}\) \leq \lambda \Tr(\hat{\rho}_{\mathrm{unt}}\hat{e})
 +\max_{\nu\leq \nu_{\mathrm{th}}} \Omega(\nu,\lambda).
\end{equation}
This inequality implies that for any $\epsilon (>0)$, the probability of violating the inequality
\begin{equation}
 \label{app-prop-main-second-ineq}
 e^{(\mathrm{ph})}_{\mathrm{unt}} \leq \lambda e_{\mathrm{unt}} + \max_{\nu\leq \nu_{\mathrm{th}}} \Omega(\nu,\lambda) +\epsilon
\end{equation}
for the measured values  decreases exponentially as the number of the untagged and detected rounds becomes larger.
 One way to prove this statement is to define a sequence regarded as a martingale and use Azuma's inequality \cite{Azuma1967}.
 
 	Next, we determine the function $\Omega(\nu,\lambda)$.
	We introduce a unitary operator $\hat{U}$ by the relation
\begin{equation}
 \hat{U} \hat{H}^{\otimes L}\ket{\ve{a}}_A\ket{k}_{B} = \hat{H}^{\otimes L}\ket{\ve{a'}}_A\ket{k}_{B},
\end{equation}
where $\ve{a}'$ is defined by $a'_l = a_l \oplus \delta_{l,k}$ for each element.
In terms of this unitary, the operators $\hat{P}^{(\nu)},\hat{e},$ and $\hat{e}^{(\mathrm{ph})}$ are transformed as
\begin{equation}
\fl
	\label{appendix-def-Upnu}
 \hat{U}^\dagger\hat{P}^{(\nu)}\hat{U} =\sum_{|\ve{a}| = \nu-1,\nu-3,\cdots} \hat{P}\(\hat{H}^{\otimes L}\ket{\ve{a}}_A\)
+\sum_{|\ve{a}|=\nu+1}\sum_{k=1}^{L}\hat{P}\(\hat{H}^{\otimes L}\ket{\ve{a}}_A\)\hat{P}\(\ket{k}_B\)\delta_{a_k,1},
\end{equation}
 \begin{equation}
	\fl
 \label{appendix-def-Ue}
 \hat{U}^\dagger\hat{e}\hat{U}=\sum_{\{k,l\}\in\mathfrak{P}} \frac{1}{L-1}\hat{P}\(\frac{1}{\sqrt{2}}\(\ket{k}_B-\ket{l}_B\)\),
 \end{equation}
\begin{eqnarray}
 \fl
 \label{appendix-def-Ueph}
  \hat{U}^\dagger\hat{e}^{(\mathrm{ph})}\hat{U} &=&\hat{e}^{(\mathrm{ph})}\nonumber\\
 &=&\frac{1}{L-1}\sum_{\ve{a}\in\mathfrak{Q}}\hat{P}\(\hat{H}^{\otimes L}\ket{\ve{a}}_A\)\sum_{k=1}^{L}\hat{P}(\ket{k}_B)\((|\ve{a}|-1)\delta_{a_k,1}+|\ve{a}|\delta_{a_k,0}\),
\end{eqnarray}
where $\mathfrak{Q}$ is defined as $\{0,1\}^{L}$.

Since all the transformed operators commute with $\hat{P}\(\hat{H}^{\otimes L}\ket{\ve{a}}_A\)$,
we have a direct-sum decomposition 
\begin{equation}
 \fl
 \label{appendix-def-Lambda}
	 \hat{U}^\dagger\hat{P}^{(\nu)}\(\hat{e}^{(\mathrm{ph})}-\lambda\hat{e}\)\hat{P}^{(\nu)}\hat{U}
	 = \sum_{|\ve{a}| = \nu+1,\nu-1,\nu-3,\cdots} \hat{P}\(\hat{H}^{\otimes L}\ket{\ve{a}}_A\) \otimes
	 \hat{\Lambda}(\nu,\lambda, \ve{a}).
\end{equation}
Due to the symmetry over the permutation of the index,
the largest eigenvalue of $\hat{\Lambda}(\nu,\lambda,\ve{a})$ depends only on $|\ve{a}|$, which we denote by $\Omega(\nu,\lambda,|\ve{a}|)$.
\Eref{appendix-def-Lambda} implies that $\Omega(\nu,\lambda)$ is equal to the maximum of $\Omega(\nu,\lambda,|\ve{a}|)$
over $|\ve{a}|=\nu+1, \nu-1, \nu-3,\cdots$.

For evaluation of $\Omega(\nu,\lambda,|\ve{a}|)$, it suffices to consider sequences in the from of $\ve{a}=(1,\cdots,1,0,\cdots,0)$.
The matrix representation of operator $\hat{\Lambda}(\nu,\lambda,\ve{a})$
is conveniently described by the following matrices.
Define $d$-dimensional matrices $M_1^{(d)}$ and $M_2^{(d,m)}$ by
 \begin{equation}
	\forall i,j, \; (M_1^{(d)})_{ij} = 1
 \end{equation}
 and
 \begin{equation}
	\forall i,j, \; (M_2^{(d,m)})_{ij} =
	 \cases{
		1& $i=j \leq m$\\
		0& otherwise\\
		 }.
 \end{equation}
We also define $\mathds{1}^{(d)}$ as the identity matrix.
 From \eref{appendix-def-Upnu}, \eref{appendix-def-Ue}, and \eref{appendix-def-Ueph}, the nonzero part of $\hat{\Lambda}(\nu,\lambda,\ve{a})$ is represented as
 \begin{equation}
	\label{appendix-Lambda-1}
		\frac{\lambda}{2(L-1)}M_1^{(L)}-\frac{1}{L-1}M_2^{(L,|\ve{a}|)}+\frac{2|\ve{a}|-L\lambda}{2(L-1)}\mathds{1}^{(L)} 
 \end{equation}
 for $|\ve{a}| = \nu-1,\nu-3,\cdots$, and 
 \begin{equation}
		\label{appendix-Lambda-2}
		\frac{\lambda}{2(L-1)}M_1^{(|\ve{a}|)} +\frac{2(|\ve{a}|-1)-\lambda L}{2(L-1)}\mathds{1}^{(|\ve{a}|)}
 \end{equation}
for $|\ve{a}|=\nu+1$.
In order to obtain their eigenvalues, we use the next proposition.
 \begin{Prop}
	\label{prop-det-bound}
	\begin{eqnarray}
	\label{prop-det-bound-result}
	 &\det\(\alpha M_1^{(d)}+\beta M_2^{(d,m)} + \gamma \mathds{1}^{(d)}\)\nonumber\\
	 =& \gamma^{d-m-1}(\gamma+\beta)^{m-1}(\gamma^2+(\beta+d\alpha)\gamma+(d-m)\alpha\beta) 
	\end{eqnarray}
 \end{Prop}
	(Proof of proposition \ref{prop-det-bound})
	The normalized vectors defined as
	\begin{equation}
	 m^{-1/2}(1,e^{\frac{2\pi i k}{m}},\cdots,e^{\frac{2\pi i k(m-1)}{m}},0,\cdots,0) \qquad (k=1,\cdots,m-1)
	\end{equation}
	and the normalized vectors  defined as
	\begin{equation}
	 \fl
	 (d-m)^{-1/2}(0,\cdots,0,1,e^{\frac{2\pi i k}{d-m}},\cdots,e^{\frac{2\pi i k(d-m-1)}{d-m}}) \qquad (k=1,\cdots,d-m-1)
	\end{equation}
	are the eigenvectors with the eigenvalues $\beta+\gamma$ and $\gamma$.
	The remaining space is spanned by the two normalized vectors
	$m^{-1/2} (1,\ldots1,0,\dots,0)$
	and 
	$(d-m)^{-1/2} (0,\ldots 0,1,\dots,1)$.
	By directly calculating the determinant, we obtain \eref{prop-det-bound-result}.
	$\qed$

	 By use of proposition \ref{prop-det-bound}, we obtain the largest eigenvalues of \eref{appendix-Lambda-1} and
	 \eref{appendix-Lambda-2} as 
\begin{eqnarray}
 \Omega_-(\nu,\lambda)&:= \max_{|\ve{a}|=\nu-1,\nu-3,\cdots}\Omega(\nu,\lambda,|\ve{a}|) = \Omega(\nu,\lambda,\nu-1)\nonumber\\
	&=\frac{\nu-1}{L-1} - \frac{L\lambda+2}{4(L-1)}\( 1- \sqrt{ 1- \frac{8(\nu-1)\lambda}{(L\lambda+2)^2}}\)
\end{eqnarray}
and
\begin{equation}
 \Omega_+(\nu,\lambda):=\Omega(\nu, \lambda,\nu+1)=\frac{\nu}{L-1} - \lambda\frac{L-1-\nu}{2(L-1)},
\end{equation}
respectively.
Thus, the function $\Omega(\nu,\lambda)$ is given by
\begin{equation}
 \Omega(\nu,\lambda) = \max \(\Omega_-(\nu,\lambda),\Omega_+(\nu,\lambda)\).
\end{equation}

From the explicit expressions of $\Omega(\nu,\lambda)$, we find that $\Omega(\nu,\lambda)$ is an increasing function for $\nu$.
It means
\begin{equation}
 \max_{\nu\leq \nu_{\mathrm{th}}} \Omega(\nu,\lambda) = \Omega(\nu_{\mathrm{th}},\lambda).
\end{equation}
%The convex function $\Omega(\nu,\lambda)$ satisfies
% \begin{eqnarray}
%	& \lim_{\lambda\rightarrow \infty} \Omega(\nu,\lambda)=
%	\lim_{\lambda\rightarrow \infty} \Omega(\nu,\lambda,\nu-1)= \frac{\nu-1}{L}\\
%	\leq&\Omega(\nu,\lambda)\\
%	\leq& \Omega(\nu,0)=\Omega(\nu,0,\nu+1)=\frac{\nu}{L-1}
	% \end{eqnarray}
Combined with \eref{prop-main-second-ineq}, it ensures that for any $\epsilon>0$, there exists $\lambda\geq 0$ such that
\begin{equation}
 \lambda e_{\mathrm{unt}} + \max_{\nu\leq\nu_{\mathrm{th}}} \Omega(\nu,\lambda) \leq F(\nu_{\mathrm{th}},e_{\mathrm{unt}}) + \epsilon.
\end{equation}
With \eref{app-prop-main-second-ineq}, we see that the probability of violating 
\begin{equation}
 e^{(\mathrm{ph})}_{\mathrm{unt}} \leq F(\nu_{\mathrm{th}},e_{\mathrm{unt}}) + 2 \epsilon 
\end{equation}
vanishes in the asymptotic limit. This statement is the precise meaning of \eref{prop-main-first-ineq}.

\end{document}